\documentclass[aps,pra,showpacs,twocolumn, longbibliography]{revtex4-1}

\usepackage{graphicx}
\usepackage{dcolumn}
\usepackage{bm}
\usepackage{color}
\usepackage{amsmath}
\usepackage{amssymb}

\newcommand{\tabincell}[2]{\begin{tabular}{@{}#1@{}}#2\end{tabular}}

\begin{document}
\title{Theoretical investigation of spectrally pure state generation from isomorphs of KDP crystal at near-infrared and telecom wavelengths}
\author{Rui-Bo Jin$^{1,4}$}
\author{Neng Cai$^{1}$}
\author{Ying Huang$^{1}$}
\author{Xiang-Ying Hao$^{1}$}
\email{xyhao.321@163.com}
\author{Shun Wang$^{1}$}
\author{Fang Li$^{1}$}
\author{Hai-Zhi Song$^{2,3}$}
\author{Qiang Zhou$^{2,5}$}
\email{zhouqiang@uestc.edu.cn}
\author{Ryosuke Shimizu$^{6}$}

\affiliation{$^{1}$ Hubei Key Laboratory of Optical Information and  Pattern Recognition, Wuhan Institute of Technology, Wuhan 430205, PR China}

\affiliation{$^{2}$ Institute of Fundamental and Frontier Science and School of Optoelectronic Science and Engineering, University of Electronic Science and Technology of China, Chengdu 610054, PR China}
\affiliation{$^{3}$Southwest Institute of Technical Physics, Chengdu 610041, PR China}
\affiliation{$^{4}$ State Key Laboratory of Quantum Optics and Quantum Optics Devices, Shanxi University, Taiyuan, 030006, PR China}
\affiliation{$^{5}$CAS Key Laboratory of Quantum Information, University of Science and Technology of China, Hefei 230026, PR China}
\affiliation{$^{6}$The University of Electro-Communications, 1-5-1 Chofugaoka, Chofu, Tokyo, Japan}
\date{\today }

\begin{abstract}
Spectrally uncorrelated biphoton state generated from the spontaneous nonlinear optical process is an important resource for quantum information. Currently such spectrally uncorrelated biphoton state can only be prepared from limited kinds of nonlinear media, thus limiting their wavelengths. In order to explore wider wavelength range, here we theoretically study the generation of spectrally uncorrelated biphoton state from 14 isomorphs of potassium dihydrogen phosphate (KDP) crystal. We find that 11 crystals from the `KDP family' still maintain similar nonlinear optical properties of KDP, such as KDP, DKDP, ADP, DADP, ADA, DADA, RDA, DRDA, RDP, DRDP and KDA, which satisfy 3 kinds of the group-velocity matching conditions for spectrally uncorrelated biphoton state generation from near-infrared to telecom wavelengths. Based on the uncorrelated biphoton state, we investigate the generation of  heralded pure-state single photon by detecting one member of the biphoton state to herald the output of the other. The purity of the heralded single photon is as high as 0.98 without using a narrow-band filter; the Hong-Ou-Mandel interference from independent sources can also achieve a visibility of  98\%. This study may provide more and better single-photon sources for quantum information processing at near-infrared and telecom wavelengths.
\end{abstract}



\maketitle

\section{Introduction}

Single photons are essential quantum information carriers for photonics quantum information processing (QIP), because one can easily manipulate, transmit and detect single photons, and then can implement many QIP protocols.  A common method to prepare single photons is based on the generation of biphoton state (usually named idler and signal photons) followed by the detection of one member of the pair (for instance the idler), which heralds the output of the other \cite{Puigibert2017, Liu2018arxiv, Joshi2018}. In this scheme a crucial step is the pair generation process, which can be achieved through spontaneous parametric down-conversion (SPDC) or spontaneous four-wave-mixing (SFWM) in a nonlinear optical medium.  The SPDC  is a widely used method for the generation of correlated pairs of photons. In the SPDC process, a higher-energy pump photon is `split' into the signal and idler biphotons, which are generally spectrally correlated due to the energy and momentum conservation laws. For many QIP applications like quantum computation \cite{Walmsley2005} or boson sampling \cite{Broome2013}, it is necessary to utilize biphotons with no spectral correlations to obtain heralded pure-state single photon, so as to achieve high distinguishability, i.e. high-visibility Hong-Ou-Mandel (HOM) interference \cite{Hong1987}, between independent single photon sources \cite{Mosley2008PRL, Jin2011, Jin2013PRA2}. A high HOM interference visibility corresponds to a high operation fidelity for several QIP protocols, such as quantum teleportation \cite{Valivarthi2016np} and measurement device independent quantum key distribution (MDI-QKD) \cite{Valivarthi2017QST, Lo2012}. Therefore it is very important to prepare spectrally uncorrelated biphotons from SPDC process.

Two methods can be applied to remove the spectral correlations between the signal and idler photons. The first one is filtering method, in which the spectral uncorrelated biphoton state is obtained by simply and directly filtering the correlated ones using narrow bandpass filters. However, this method may not only severely decrease the brightness of the photon source, but also introduce extra noises \cite{Mosley2008PRL, Meyer-Scott2017}.
The second one is quantum state engineering method \cite{Grice2001,  Konig2004, Edamatsu2011}, in which, by engineering the group-velocity-matched (GVM) condition, spectrally uncorrelated biphoton state can be intrinsically generated in specific crystals at several fixed wavelengths. For example, the KDP crystal at 830 nm has a maximal purity of 0.97 \cite{Mosley2008PRL, Jin2011, Jin2013PRA2}; the  $\beta$-barium borate (BBO) crystal at 1514 nm can achieve a maximal purity of 0.82 \cite{Grice2001, Lutz2013OL, Lutz2014, TingyuLi2018OE}; the periodically poled KTP crystal (PPKTP) at 1584 nm has a maximal purity of 0.82 \cite{Evans2010, Gerrits2011, Eckstein2011, Jin2013PRA, Zhou2013, Jin2014OE, Weston2016, Wang2014, Bruno2014OE, Jin2016SR, Jin2018Optica, Greganti2018, Zhou2018PRL, Jin2018PRAppl, Jin2018OE, Meyer-Scott2018arXiv, Terashima2018} and this high purity can be kept when the wavelength is tuned from 1400nm to 1700 nm \cite{Jin2013OE}.  The purity for PPKTP crystals can be further improved from 0.82 to near 1 using the custom poling crystal \cite{Branczyk2011, Dixon2013, Dosseva2016, Tambasco2016, Chen2017}.  Such GVM condition can also be satisfied in the SFWM process in optical fiber with birefringence \cite{Smith2009, Fang2013}.

As described above, the spectrally uncorrelated biphoton state is an important but rare resource. At present, researchers have only used limited kinds of crystals (KDP, BBO, PPKTP, etc) to produce such biphoton state in a limited wavelength range. It should be valuable to find more crystals to prepare spectrally uncorrelated biphoton state, then to prepare heralded pure-state single photon in a wider wavelength range. Recently, we have found that four isomorphs of PPKTP (PPRTP, PPKTA, PPRTA, PPCTA) can retain the properties of PPKTP, i.e, these isomorphs satisfy the GVM condition and can prepare spectrally uncorrelated biphoton state in the range of 1300 nm to 2100 nm \cite{Jin2016PRAppl, Laudenbach2017, Jin2019JOLT}.

In this work, inspired by the previous studies, we investigate the possibility of the generation of spectrally uncorrelated biphoton state from 14 isomorphs from the `KDP family'.~This article is organized as follows. First, the introductory part provides the background and  motivation of this research. Second, the theoretical part introduces the basic characteristics of KDP isomorphous crystals, and the principle of spectrally uncorrelated biphoton state generation based on GVM conditions. Third, the simulation part shows the  wavelength degenerate case, nondegenerate case and the Hong-Ou-Mandel (HOM) interference patterns with independent heralded spectrally-pure single-photon state. Fourth, we provide fruitful discussions on the simulation results in the discussion section. Finally, we summarize the paper in the conclusion section.

\section{Theory }
\subsection{The basic characteristics of the `KDP family' }
The general form of the isomorphs of KDP can be written as $M$D$X$ or D$M$D$X$ with {$M=K, Rb, Cs, NH_{4}$} and {$X=P, As$}. Here, $K, Rb, Cs$ are alkali metal elements of group $IA$ in the periodic table; $P$ and $As$ are elements of  group $VA$ in the periodic table; and the meaning of the first letter D in D$M$D$X$ is to replace the protium element in $M$D$X$ with deuterium. These crystals include KDP ($KH_2PO_4$), DKDP ($KD_2PO_4$), ADP ($NH_4H_2PO_4$), DADP ($ND_4D_2PO_4$), ADA ($NH_4H_2AsO_4$), DADA ($ND_4D_2AsO_4$),  RDA ($RbH_2AsO_4$), DRDA ($RbD_2AsO_4$), RDP ($RbH_2PO_4$),  DRDP ($RbD_2PO_4$), KDA ($KH_2AsO_4$), DKDA ($KD_2AsO_4$), CDA ($CsH_2AsO_4$) and DCDA ($CsD_2AsO_4$)~\cite{Dmitriev1999, Nikogosyan2005}. All these 14 crystals belong to the `KDP  family' and have the similar properties as the KDP crystal: (1)  all of them are  negative uniaxial crystal ($n_o>n_e$, point group $\overline4$2m);  (2)  all of them  have high transmission rate from ultraviolet to infrared; (3)  all of them have high optical damage threshold; (4)  all of them are widely used in lasers, optical switches, electro-optic modulators, etc.

\subsection{The principle of spectrally uncorrelated biphoton state generation based on GVM condition}

The biphoton state $\vert\psi\rangle$ generated from the process of SPDC can be written as
\begin{equation}\label{eq1}
\vert\psi\rangle=\int_0^\infty\int_0^\infty\,\mathrm{d}\omega_s\,\mathrm{d}\omega_if(\omega_s,\omega_i)\hat a_s^\dag(\omega_s)\hat a_i^\dag(\omega_i)\vert0\rangle\vert0\rangle,
\end{equation}
where $\omega$ is the angular frequency; the subscripts $s$ and $i$ indicate the signal and idler photon; $\hat a^\dag$ is the creation operator. The joint spectral amplitude (JSA) $f(\omega_s,\omega_i)$ is the product of the pump envelope function (PEF) $\alpha(\omega_s,\omega_i)$ and the phase matching function (PMF) $\phi(\omega_s,\omega_i)$, i.e.,
\begin{equation}\label{eq2}
f(\omega_s,\omega_i) = \alpha(\omega_s,\omega_i) \times \phi(\omega_s,\omega_i).
\end{equation}
A PEF with a Gaussian-distribution can be written as \cite{Mosley2008NJP},
\begin{equation}\label{eq20}
\alpha(\omega_s, \omega_i)=\exp[-\frac{1}{2}\left(\frac{\omega_s+\omega_i-\omega_p}{\sigma_p}\right)^2],
\end{equation}
where $\sigma_p$ is the bandwidth of the pump.
Using wavelengths as the variables, the PEF can be rewritten as
\begin{equation}\label{eq21}
\alpha(\lambda_s, \lambda_i)= \exp \left\{  -\frac{1}{2} \left( \frac{{1/\lambda _s  + 1/\lambda _i  - 1/(\lambda _0 /2)}}{{\Delta \lambda /[(\lambda _0 /2)^2  - (\Delta \lambda /2)^2 ]}}  \right) \right\},
\end{equation}
where  $\lambda _0 /2 $ is the central wavelength of the pump; $\Delta \lambda$ is the bandwidth in wavelength and $\sigma _p$ is,
\begin{equation}\label{eq31}
\sigma _p  = \frac{{2\pi c\Delta \lambda }}{{(\lambda _0 /2)^2  - (\Delta \lambda /2)^2 }},
\end{equation}
where $c$ is the speed of light.
By assuming a flat phase distribution, the PMF function can be written as \cite{Mosley2008NJP}
\begin{equation}\label{eq22}
\phi(\omega_s,\omega_i)=\operatorname{sinc}\left(\frac{\Delta kL}{2}\right),
\end{equation}
where $L$ is the length of crystal, $\Delta k=k_p-k_i-k_s$ and $k=\frac{2 \pi n (\lambda, \varphi)}{\lambda}$ is the wave vector.  For ordinary ray (o-ray), the refractive index $n_o (\lambda )$  is a function of wavelength $\lambda$.  While for extraordinary ray (e-ray), the refractive index $n_e (\lambda, \varphi)$  is a function of $\varphi$ and wavelength $\lambda$. $\varphi$ is the angle between the optical axis of the crystal and the direction of the pump laser.
In this study, we set the pump and idler are e-rays, and the signal is o-ray. Then, $\Delta k$ can be rewritten as
\begin{equation}\label{DeltaK}
\Delta k = 2\pi \left[\frac{{n_e (\lambda _p ,\varphi )}}{{\lambda _p }} - \frac{{n_o (\lambda _s )}}{{\lambda _s }} - \frac{{n_e (\lambda _i ,\varphi )}}{{\lambda _i }}\right].
\end{equation}

While the phase mismatch is zero, i.e. $\Delta k=0$,
the phase is completely matched, therefore, it is called the phase matching condition. The angle between the positive direction of horizontal axis and the ridge direction of the PEF is always 135$^ \circ$ \cite{Jin2013OE}. In contrast, the angle $\theta$ between the horizontal axis and the ridge of the PMF is determined by the following equation \cite{Jin2013OE}
\begin{equation}\label{eq4}
\tan\theta=-\left( \frac{V_{g,p}^{-1}(\omega_p)-V_{g,s}^{-1}(\omega_s)}{V_{g,p}^{-1}(\omega_p)-V_{g,i}^{-1}(\omega_i)} \right),
\end{equation}
where  $V_{g,\mu}=\frac{d\omega}{dk_\mu(\omega)}=\frac{1}{k_\mu^\prime(\omega)},(\mu=p, s, i)$ is the group velocity of the pump, the signal and the idler.
In other words,  the shape of the PMF is determined by the GVM condition.
We consider three kinds of GVM condition in this work  \cite{Edamatsu2011}.
The first GVM condition (GVM$_1$)  is,
\begin{equation}\label{gvm1}
V_{g,p}^{-1}(\omega_p)=V_{g,s}^{-1}(\omega_s).
\end{equation}
The second GVM condition (GVM$_2$) is,
\begin{equation}\label{gvm2}
V_{g,p}^{-1}(\omega_p)=V_{g,i}^{-1}(\omega_i).
\end{equation}
The third GVM condition (GVM$_3$) is,
\begin{equation}\label{gvm3}
2V_{g,p}^{-1}(\omega_p)=V_{g,s}^{-1}(\omega_i)+V_{g,s}^{-1}(\omega_s).
\end{equation}
When the GVM$_1$, GVM$_2$ or GVM$_3$ conditions are satisfied, $\varphi = 0^{\circ}$, $90^{\circ}$ or $45^{\circ}$, respectively.
With these three kinds of GVM conditions, it is possible to prepare spectrally uncorrelated biphotons. In next section, we calculate the parameters for GVM conditions in detail \cite{Supplement2019}.
%
%

%
%
\begin{figure}[bp]
\centering\includegraphics[width=6cm]{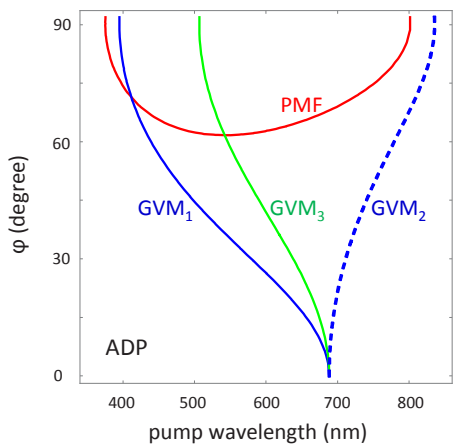}
\caption{The phase matched function (PMF) and group velocity matched functions (GVM$_1$, GVM$_2$ and GVM$_3$) for different pump wavelengths $\lambda$ and phase matched angle $\varphi$ for ADP crystal. In this calculation, we consider the Type-II  ($e \to o + e$) condition with collinear and wavelength-degenerated ($2\lambda_p=\lambda_s=\lambda_i$) configuration.
 } \label{pmfgvm123}
\end{figure}
\begin{table*}[tbp]
\centering
\begin{tabular}{ c|c c c}
\hline \hline
Name&GVM$_1$  (purity $ \approx $ 0.97)& GVM$_2$  (purity $ \approx $ 0.97) & GVM$_3$  (purity $ \approx $ 0.82) \\
\hline
KDP& \tabincell{l}{ $\lambda_p$=415 nm, $\lambda_{s,i}$=830 nm \\  $\varphi$ =67.7$^\circ$,  $d_\textrm{eff}$=0.28 pm/V}
     & not satisfied
      & \tabincell{l}{$\lambda_p$=551 nm, $\lambda_{s,i}$=1102 nm \\  $\varphi$ =59.0$^\circ$,  $d_\textrm{eff}$=0.33 pm/V}  \\
\hline
DKDP& \tabincell{l}{$\lambda_p$=476 nm, $\lambda_{s,i}$=952 nm \\  $\varphi$ =57.6$^\circ$,  $d_\textrm{eff}$=0.22 pm/V}
        &  \tabincell{l}{$\lambda_p$=915 nm, $\lambda_{s,i}$=1830 nm \\  $\varphi$ =62.9$^\circ$,  $d_\textrm{eff}$=unknown}
        &  \tabincell{l}{$\lambda_p$=626 nm, $\lambda_{s,i}$=1252 nm \\  $\varphi$ =51.7$^\circ$,  $d_\textrm{eff}$=0.34 pm/V}  \\
\hline
ADP& \tabincell{l}{$\lambda_p$=411 nm, $\lambda_{s,i}$=822 nm \\  $\varphi$ =71.2$^\circ$,  $d_\textrm{eff}$=0.34 pm/V}
      &   not satisfied
      &  \tabincell{l}{$\lambda_p$=541 nm, $\lambda_{s,i}$=1082 nm \\  $\varphi$ =61.5$^\circ$,  $d_\textrm{eff}$=0.44 pm/V}  \\
\hline
DADP& \tabincell{l}{$\lambda_p$=464 nm, $\lambda_{s,i}$=978 nm \\  $\varphi$ =59.6$^\circ$,  $d_\textrm{eff}$=0.45 pm/V}
      &  \tabincell{l}{$\lambda_p$=869 nm, $\lambda_{s,i}$=1738 nm \\  $\varphi$ =64.2$^\circ$,  $d_\textrm{eff}$=unknown}
      &  \tabincell{l}{$\lambda_p$=609 nm, $\lambda_{s,i}$=1218 nm \\  $\varphi$ =53.2$^\circ$,  $d_\textrm{eff}$=0.48 pm/V}  \\
\hline
ADA& \tabincell{l}{$\lambda_p$=461 nm, $\lambda_{s,i}$=922 nm \\  $\varphi$ =69.8$^\circ$,  $d_\textrm{eff}$=0.26 pm/V}
      &  not satisfied
      &  \tabincell{l}{$\lambda_p$=605 nm, $\lambda_{s,i}$=1210 nm \\  $\varphi$ =60.6$^\circ$,  $d_\textrm{eff}$= unknown}  \\
\hline
DADA& \tabincell{l}{$\lambda_p$=522 nm, $\lambda_{s,i}$=1044 nm \\  $\varphi$ =57.0$^\circ$,  $d_\textrm{eff}$=0.33 pm/V}
      &  not satisfied
      &  \tabincell{l}{$\lambda_p$=741 nm, $\lambda_{s,i}$=1482 nm \\  $\varphi$ =49.5$^\circ$,  $d_\textrm{eff}$= unknown}  \\
\hline
RDA& not satisfied
      &  not satisfied
      &  \tabincell{l}{$\lambda_p$=648 nm, $\lambda_{s,i}$=1296 nm \\  $\varphi$ =72.5$^\circ$,  $d_\textrm{eff}$=0.21 pm/V}  \\
\hline
DRDA& \tabincell{l}{$\lambda_p$=546 nm, $\lambda_{s,i}$=1092 nm \\  $\varphi$ =74.1$^\circ$,  $d_\textrm{eff}$=0.20 pm/V}
      &  not satisfied
      &  \tabincell{l}{$\lambda_p$=727 nm, $\lambda_{s,i}$=1454 nm \\  $\varphi$ =62.8$^\circ$,  $d_\textrm{eff}$=0.29 pm/V}  \\
\hline
RDP& not satisfied
      & not satisfied
      &  \tabincell{l}{$\lambda_p$=578 nm, $\lambda_{s,i}$=1156 nm \\  $\varphi$ =82.1$^\circ$,  $d_\textrm{eff}$=0.10 pm/V}  \\
\hline
DRDP& not satisfied
      &  not satisfied
      &  \tabincell{l}{$\lambda_p$=639 nm, $\lambda_{s,i}$=1278 nm \\  $\varphi$ =70.0$^\circ$,  $d_\textrm{eff}$=0.21 pm/V}  \\
\hline
KDA& \tabincell{l}{$\lambda_p$=467 nm, $\lambda_{s,i}$=934 nm \\  $\varphi$ =68.7$^\circ$,  $d_\textrm{eff}$=0.40 pm/V}
      &  not satisfied
      &  not satisfied  \\
\hline \hline
\end{tabular}
\caption{\label{table1} Three kinds of GVM conditions for eleven crystals. $\lambda_{p(s,i)}$ is the GVM wavelength for the pump (signal, idler). $\varphi$ is the phase matched angle and $d_\textrm{eff}$ is the effective nonlinear coefficient.
The $d_\textrm{eff}$ values are taken from the SNLO $v66$ software package, developed by AS-Photonics, LLC \cite{SNLO66}. Some of the $d_\textrm{eff}$ values are not available in this software. The Sellmeier equations are obtained from Ref. \cite{Dmitriev1999} and \cite{Nikogosyan2005}.
Note that CDA and DCDA don't satisfy the GVM condition;  the Sellmeier equation of DKDA has not been reported yet.
 }
\end{table*}
\begin{figure*}[tbp]
\centering\includegraphics[width=13cm]{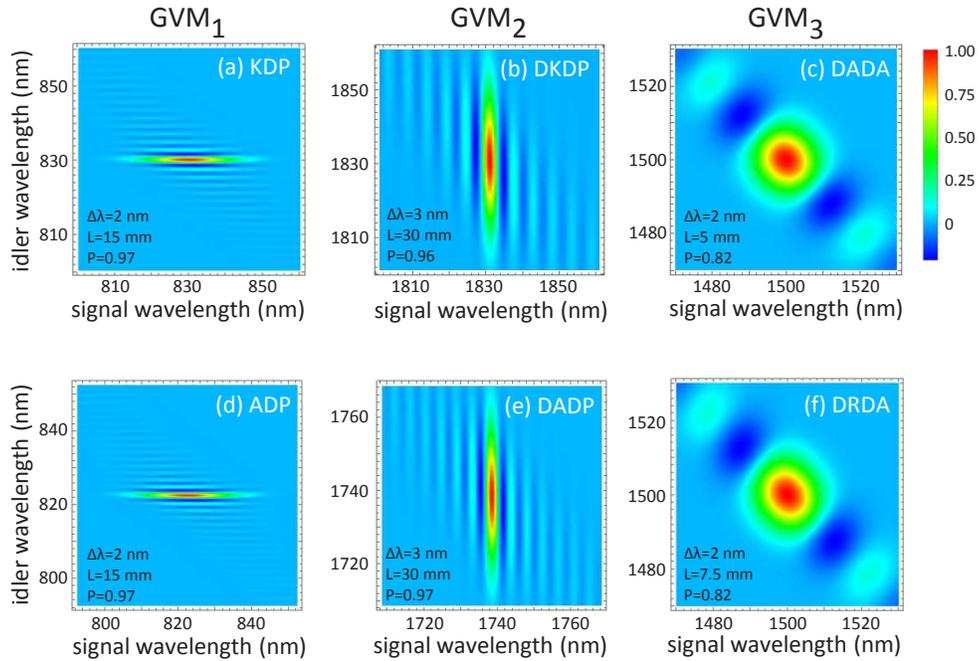}
\caption{The JSA of biphotons generated from the isomorphs of KDP. The bandwidth  $\Delta \lambda$, the crystal length $L$ and the purity $P$ are listed in each figure. Figures (a, d) in the first column are for GVM$_1$ condition. Figures  the second and third columns are for  GVM$_2$  and GVM$_3$ conditions respectively.
 } \label{jsi6}
\end{figure*}
\section{Simulation}

\begin{table*}[tbp]
\centering
\begin{tabular}{ c|l l l}
\hline \hline
Name&GVM$_1$ condition \\
\hline
RDA&    $\lambda_p$=520 nm, $\lambda_s$=764 nm,  $\lambda_i$=1630 nm,  $\varphi$ =56.0$^\circ$,  $d_\textrm{eff}$=unknown  \\
\hline
DRDP&    $\lambda_p$=500 nm, $\lambda_s$=744 nm,  $\lambda_i$=1526 nm,  $\varphi$ =56.0$^\circ$,  $d_\textrm{eff}$=unknown\\
\hline
KDA&    $\lambda_p$=520 nm, $\lambda_s$=787 nm,  $\lambda_i$=1531 nm,  $\varphi$ =45.1$^\circ$,  $d_\textrm{eff}$=0.52 pm/V  \\
\hline \hline
\end{tabular}
\caption{\label{table2} The parameters of three crystals. $\varphi$ is the phase matched angle and $d_\textrm{eff}$ is the effective nonlinear coefficient.
The $d_\textrm{eff}$ values are taken from the SNLO $v66$ software package, developed by AS-Photonics, LLC \cite{SNLO66}. Some of the $d_\textrm{eff}$ values are not available in this software.
 }
\end{table*}

\begin{figure}[tbp]
\centering\includegraphics[width=8.5cm]{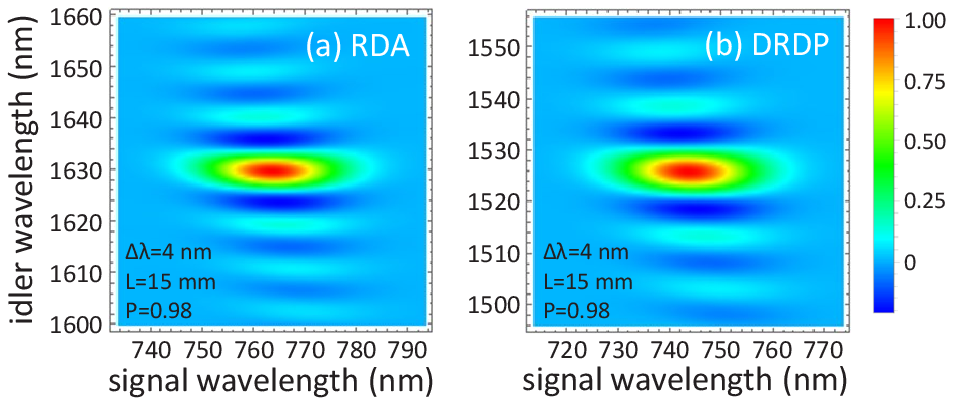}
\caption{ The JSA of biphotons generated from  RDA and DRDP. The center wavelengths are set to be GVM wavelength in Tab. \ref{table2} for each crystal. The  pump bandwidth $\Delta \lambda=4 $ nm and the crystal length $L$=15 mm.
 } \label{nondege}
\end{figure}

\subsection{Wavelength degenerate case}

First, we consider the Type-II ($e \to o + e$)  SPDC with collinear and wavelength-degenerated ($2\lambda_p=\lambda_s=\lambda_i$) configurations.
According to GVM conditions,  the GVM wavelength $\lambda_{p(s, i)}$ and the corresponding phase matched angle  $\varphi$ can be calculated.
As an example, Fig. \ref{pmfgvm123} shows the PMF and GVM$_{1(2,3)}$ conditions for different wavelengths and phase matched angles for ADP crystal. The cross points  correspond the case PMF and GVM conditions are simultaneously satisfied.
Following the similar way, we can also calculate the GVM conditions for other crystals. Table\,\ref{table1} lists the details of the three kinds of GVM conditions for 11 crystals from the `KDP family'. In Tab.\,\ref{table1}, the down converted photons have a wavelength range from 822 to 1830 nm, and the corresponding pump wavelength range is from 411 to 915 nm.
The purity,  a parameter to characterize the spectral correlations of the biphotons, can be numerically calculated by conducting a Schmidt decomposition on the JSA, i.e. $f(\omega_1,\omega_2)$ -- see more details in Ref. \cite{Jin2013OE}.
The maximal purities are around 0.97 for photon pairs generated from GVM$_1$ and GVM$_2$ conditions, and 0.82 for the case of GVM$_3$ condition. Our analysis indicates these `KDP family' crystals can be used to generate spectrally uncorrelated biphoton state.

In order to investigate more details of the GVM conditions, we choose six typical configurations from Tab.\,\ref{table1}  and plot their JSAs. The results are shown in  Fig. \ref{jsi6}(a-f).
Figure \ref{jsi6}(a, d) in the first column show the  JSAs of GVM$_1$ condition, which has a long-strip shape in the horizontal direction. Here, we set the crystal length $L$ = 15 mm and the pump bandwidth $\Delta \lambda$ = 2 nm. The degenerated wavelength for KDP and ADP  are 830 nm and 822 nm, respectively. These wavelengths are the typical near-infrared wavelength where silicon APD has high detecting efficiency.
Figures \ref{jsi6}(b, e) in the second  column show the  JSAs of GVM$_2$ condition, which has a long-strip shape in the vertical direction. Here, we set the crystal length $L$ = 30 mm and the  pump bandwidth  $\Delta \lambda$ = 3 nm. The purity is 0.96 for DKDP and 0.97 for DADP. The degenerated wavelength for DKDP and DADP  are 1830 nm and 1738 nm, respectively.
Figures \ref{jsi6}(c, f) in the third column show the JSAs of GVM$_3$ condition, which has a near-round shape in the center and with side-lobes in the anti-diagonal directions.
For GVM$_3$  condition, the tunable range in wavelength is very wide. So, without degradation of the purity, we can shift the wavelength for DADA from 1482 nm (GVM$_3$ wavelength  in Tab.\,\ref{table1}) to 1500 nm (in Fig. \ref{jsi6}(c)), which is in the S-band in telecom wavelengths.
The wavelength of DRDA crystal  can also be shifted from 1454 nm (GVM$_3$ wavelength  in Tab.\,\ref{table1}) to 1500 nm in Fig. \ref{jsi6}(f).
This wide tunability is  as the same as the case of PPKTP crystal, which has a wavelength tunable range of more than 200 nm, while keep the purity around 0.82 \cite{Jin2013OE}.

All the JSAs in Fig. \ref{jsi6}(a-f) and Tab.\,\ref{table1} have high spectral purities,  indicating that the isomorphs of KDP retain the KDP-like properties. Furthermore, it also can be seen that the wavelengths, at which spectrally uncorrelated biphoton state can be generated, range from 822 nm to 1830 nm with different crystals. This shows that isomorphs of KDP can extend the wavelength range for spectrally uncorrelated biphoton state generation.

\subsection{Wavelength  nondegenerate  case}
In the nondegenerate case, we consider the wavelength of the pump photons locate at 400 $\sim$ 540 nm, which are the wavelengths for commercial-available, low-cost laser diodes; One of the down converted photons is at 500 $\sim$ 900 nm wavelength range, where the silicon avalanche photodiode (APD) has good detection  performance; The other  down converted  photon is at telecom wavelength around 1550 nm, for low-loss long-distance transmission in optical fibers. The typical configurations include 405 nm $\to $ 548 nm + 1550 nm,  450 nm $\to $ 635 nm + 1550 nm,  520 nm $\to $ 783 nm + 1550 nm and 530 nm $\to $ 806 nm + 1550 nm. These configurations are connecting the near-infrared wavelength and the telecom wavelength \cite{Kaneda2016, Saglamyurek2011nature}.

Calculation results show that 3 out of the 11 crystals can satisfy the GVM$_1$ condition in a type-II ($e \to o + e$)  SPDC with collinear and wavelength-nondegenerated ($\lambda_s  \ne \lambda_i$) configurations.
Table \,\ref{table2} lists the GVM$_1$ conditions for RDA, DRDP and KDA,  and the JSAs under these conditions are shown in Fig. \ref{nondege} and Fig. \ref{homi}(a). All the JSAs in Fig. \ref{nondege} and Fig. \ref{homi}(a) have high purities of 0.98.

\subsection{HOM interference between independent sources}
Based on the generation of spectrally uncorrelated biphoton state, we can obtain heralded pure-state single photons. For example, using one KDA crystal, we can produce a signal ($s_{1}$) and idler ($i_{1}$) photon pair. We keep $s_{1}$  as heralded single photon, while detect $i_{1}$  by a single photon detector (SPD), and use the output of the SPD as heralding signal of the output of $s_{1}$. Next, we verify the indistinguishability of generated heralded single photons, which is realized by checking the Hong-Ou-Mandel (HOM) interference \cite{Hong1987} with two independent heralded single photon sources, with a typical experimental setup shown in Refs. \cite{Mosley2008PRL, Jin2013PRA}.
In this interference, two signals $s_{1}$ and $s_{2}$ are sent to a beamsplitter for interference, and two idlers $i_{1}$ and $i_{2}$ are detected by SPDs for heralding the signals.
The four-fold coincidence counts $P$ as a function of $\tau$ can be described by Eq.\,(\ref{eq7})~\cite{Jin2015OE},
\begin{equation}\label{eq7}
\begin{split}
P (\tau )  = & \frac{1}{4}  \int_0^\infty \int_0^\infty \int_0^\infty \int_0^\infty d\omega _{s_1} d\omega _{s_2} d\omega _{i_1} d\omega _{i_2}  \\ & {\rm{|}}f_1 (\omega _{s_1} ,\omega _{i_1} )f_2 (\omega _{s_2} ,\omega _{i_2} )- \\ & f_1 (\omega _{s_2} ,\omega _{i1} )f_2 (\omega _{s_1} ,\omega _{i_2} )e^{ - i(\omega _{s_2}  - \omega _{s_1} )\tau } {\rm{|}}^{\rm{2}},
\end{split}
\end{equation}
where $f_1$ and $f_2$ are the JSAs from the first and the second crystals.

\begin{figure*}[tbp]
\centering\includegraphics[width=13cm]{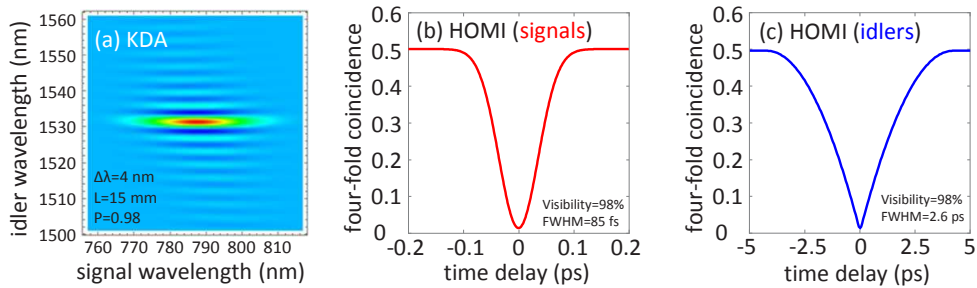}
\caption{ (a) The JSA of the biphoton generated from KDA; (b) HOM interference curve with two heralded signal photons from two independent DKDP sources; (c) HOM interference curve with two heralded idler photons.
 } \label{homi}
\end{figure*}
Figure \ref{homi}(a) is the JSA of biphoton state from KDA crystal, with crystal length $L$=15 mm and a pump bandwidth of $\Delta \lambda=2 $ nm. Under this condition the spectral purity is as high as 0.98.
Figure \ref{homi}(b, c) are the HOM interference curves between two heralded signals or two heralded idlers. Without using any narrow bandpass filters, the visibility can achieve 98\%. Please note that the visibility equals to the purity for the ideal case in the HOM interference between independent sources \cite{Mosley2008PRL, Takesue2012, Jin2013PRA}.

\section{Discussions}

A purity of 98\% is a good metric in our investigation, but it may require even higher for practical applications, e.g., for the cascaded operations, or for high-fidelity quantum information processing. Purity higher than 98\% may be realized by inserting a coarse bandpass filter, or adopting waveguide structure. Especially, utilizing the geometry dispersion in a waveguide structure also has other the advantages of single spatial mode, high brightness, on-chip integration, etc. For instance, in the pioneering work from Justin B. Spring et al., they demonstrated a near identical pure photon source based on a silica chip \cite{Spring2017Optica}.

For experimental realization, the working regime required by our calculation can be straightforwardly realized. For example, the center wavelength of the pump can be accurately controlled by utilizing a wavelength-tunable Ti: Sapphire laser; The bandwidth of the pump can be manipulated by using a bandpass filter; The crystals with the required conditions (including the crystal length and the cutting angle) are also commercial available.

In our work, we have also tried CDA and DCDA,  but these two crystals do not  satisfy the GVM condition and therefore cannot be used for spectrally uncorrelated biphoton state preparation. One possible reason for this situation is that these crystals may have some special properties, therefore they do not maintain the properties of KDP. It also could be the reason that the inaccurate test of the Sellmeier equations for these crystals in previous experiments.
In addition, the Sellmeier equation of DKDA crystal has not been reported.  We also would like to point out that our theoretical expectations can be easily verified experimentally by measuring the JSA of the generated biphoton states \cite{Mosley2008PRL, Jin2011}.

Our current work only discusses some pure-element crystals. It is possible to mix these elements to obtain mixed crystals, for example, the  KADP crystal~\cite{Ren2008}. The mixed crystal may also maintain the properties of the KDP and it is worth making further investigations in the future. If deuterium element is replaced by Tritium, we may obtain a series of new isomorphic crystals T$M$D$X$ ($M=K, Rb, Cs, NH_{4}$; $X=P, As$), which may have novel nonlinear optical properties in the GVM wavelengths, nonlinear coefficients, damage threshold, etc. It is also possible to mix three isotopes of hydrogen, i.e., protium, deuterium and tritium, so as to make mixed crystal for better performance in nonlinear optics.

In this paper, we only consider the collinear matched SPDC. For non-collinear matching, the satisfied wavelength range can be further expanded, and one can greatly increase the wavelength range of the spectrally uncorrelated biphoton state, thus for the heralded pure-state single photon source.

For future applications, the biphotons in Fig. \ref{jsi6}(a, d) can be applied for near-infrared wavelength;  Fig. \ref{jsi6}(c, f) is useful for telecom wavelength; Fig. \ref{nondege} and Fig. \ref{homi}(a) are good candidates for connecting the near-infrared wavelength and telecom wavelength, which may have great potential in quantum networks~\cite{Sangouard2011rmp}.
Figures \ref{jsi6}(a, b, d, e), \ref{nondege} and \ref{homi}(a) can be used for interference between two signals from two independent SPDC sources. Figure~\ref{jsi6}(c, f) is useful for interference between the signal and idler photons from one SPDC source.
The highly pure single photon sources at telecom wavelengths are very important for practical applications which require long-distance transmission in low-loss and low-cost optical fibers, for example, for quantum key distribution \cite{Zhangqiang2018oe}, quantum digital signature \cite{Collins2017SR}, quantum direct communication \cite{Hu2016lsa}, quantum teleportation \cite{Jin2015SRswap}, quantum networks \cite{Wengerowsky2018}, etc.

\begin{figure*}[tbp]
\centering\includegraphics[width=16cm]{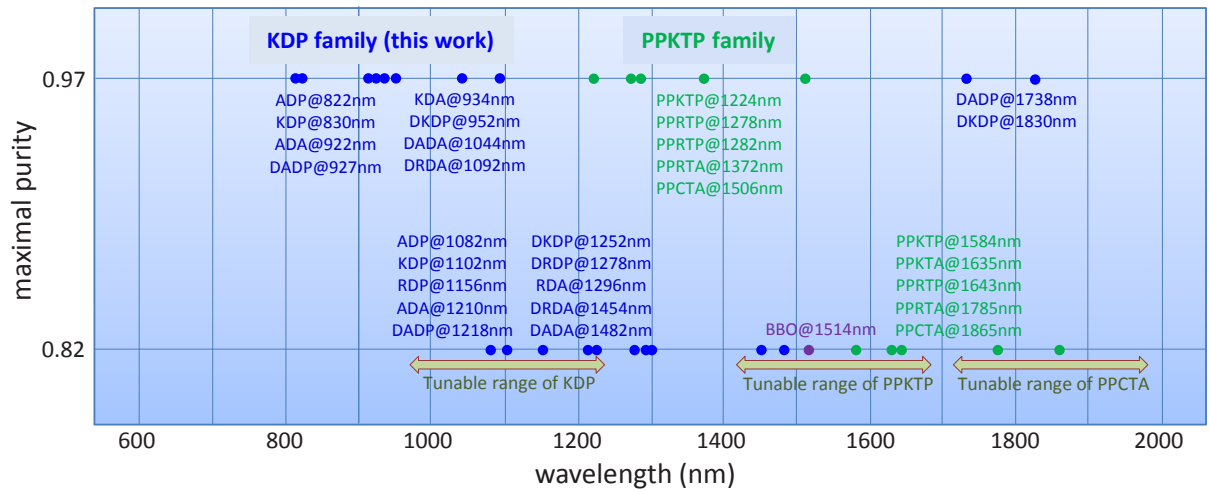}
\caption{ The wavelength versus achievable maximal spectral purity for different GVM-matched crystals. The `KDP family' (blue points) can achieve purity of  0.97 from 822 nm to 1830 nm, and 0.82 from 1082 nm to 1482 nm. The BBO crystal  (purple point) can achieve purity of  0.82 at 1514 nm \cite{Grice2001, Lutz2013OL, Lutz2014}. The `PPKTP family' (green points) can achieve purity of 0.97 from 1224 nm to 1506 nm, and 0.82  from 1584 nm to 1865 nm.  For, GVM$_3$ condition, the  purity of 0.82 can be kept when the wavelength is tuned for more than 200 nm.}
 \label{allcrystal}
\end{figure*}

Although the nonlinear coefficients of `KDP family' are generally lower than that of BBO and PPKTP, the KDP isomorph crystals have their own advantages and can not be easily replaced.
For instance, the KDP family can achieve the maximal spectral purity of ~0.98, much higher than the value of ~0.82 by the BBO \cite{Grice2001, Lutz2013OL, Lutz2014},  as summarized in Fig. \ref{allcrystal}. The size of DKDP crystals can achieve 900 mm \cite{DeYoreo2002}, much larger than the typical size of 1$\times$2$\times$30 mm$^{3}$ for PPKTP \cite{Jin2013OE}. The large size, fast growth rate, and high  damage thresholds make KDP useful for laser nuclear fusion in the National Ignition Facility \cite{DeYoreo2002}. Historically, the first SPDC experiment was realized in the ADP crystal \cite{Burnham1970}, and later several important experiments were also realized in KDP crystals \cite{Mosley2008PRL, Mosley2008NJP, Jin2011}. We expect that our investigated 11 kinds of isomorphs from the `KDP family' will also have wide applications in the future.

\section{Conclusion}
By theoretical calculation and numerical simulation, we have investigated the preparation of spectrally uncorrelated biphoton state and heralded pure-state single photon source from 14 isomorphs of the `KDP family'. It was  shown that 11 kinds of crystals, namely KDP, DKDP, ADP, DADP, ADA, DADA, RDA,  DRDA, RDP, DRDP and KDA still maintain the characteristics of KDP. For instance, they can satisfy 3 kinds of GVM condition from near infrared to telecom bands; can prepare spectrally uncorrelated biphoton state with purity as high as 0.98; and can achieve a visibility of 98\% in the HOM interference between two independent heralded pure-state single photon sources. Our work will provide good single photon sources for photonics quantum information from near-infrared to telecom wavelength.


\section*{Acknowledgments}
 This work is  partially supported by the National Key R$\&$D Program of China (Grant No. 2018YFA0307400), the National Natural Science Foundations of China (Grant Nos.91836102, 11704290, 61775025, 61405030, 11104210), and by the Program of State Key Laboratory of Quantum Optics and Quantum Optics Devices (No: KF201813), the fund from the Educational Department of Hubei Province, China (Grant No. D20161504) and the open research fund of Key Lab of Broadband Wireless Communication and Sensor Network Technology (Nanjing University of Posts and Telecommunications), Ministry of Education (Grant No. JZNY201709). We thank Guo-Qun Chen for helpful discussions.

\end{document}